\documentstyle[aps,preprint,prb
]{revtex} 
\author{F.N. Braun}
\address{Department of Chemistry, Heriot-Watt University,
Edinburgh EH14 4AS, United Kingdom}
\title{Hydrophobic interaction in the liquid phases of globular 
protein solutions: structure factor parameters}
\begin{document}
\maketitle
\begin{abstract}
We develop a simple correspondence between 
hydrophobic surface topology of globular proteins
and and an effective protein-protein adhesiveness  
parameter of the Baxter type. 
We discuss within this framework analytical interpretation
of the structure factor governing static light scattering.
\end{abstract}
\pacs{}

Hydrophobic amino-acid residues exposed at the surface of globular proteins 
are able to avoid the aqueous environment to an extent by burying
themselves in globule-globule contacts.
There exists in this sense an effective globule-globule adhesive potential,
which, from a thermal perspective, mirrors the strong entropic 
character\cite{Ben,Chandler} of hydrophobicity,
driving a general tendency of proteins to cluster with increasing temperature.

In principle, the effect is experimentally accessible
via measurement of scattered light intensity at low wavenumber $qd<<1$, 
\[
I\sim cS_0(c), 
\]
where $S_0$ denotes the $q=0$ limit of the structure factor, $c$ is the 
protein concentration, and $d$ characterises the protein globule lengthscale.

Our objective here is to supplement the usual DLVO interpretation of $S_0$
adopted, for example, in recent scattering studies 
of lysozyme\cite{Muschol} and $\beta$-lactoglobulin,\cite{Piazza}
with a semi-empirical incorporation of the hydrophobic adhesive mechanism.
We do this in the following by first establishing an effective attractive 
well interaction between proteins, having an explicit temperature dependence 
derived from thermodynamic data, and then
mapping the well interaction onto the adhesiveness parameter
of Baxter, this latter being very amenable to liquid state analysis.
\medskip

The starting point is a formulation of 
`hydrophobic surface tension' $\gamma$ according 
to a typical transfer free energy $\Delta F\sim$ 2 kcal/mol of a single 
nonpolar amino-acid residue between water and a hydrophobic environment, i.e.,
\cite{Vallone}
\begin{equation}
\gamma= g \Delta F/a^2
\end{equation}
where $g$ is some geometrical factor less than one, and $a\sim 1$nm 
is a typical residue dimension.

An empirical temperature dependence $\gamma(T)$ follows via this 
expression by extrapolating from room temperature thermodynamic data 
available for $\Delta F$, 
\[
\Delta F(T)=\Delta H^0+\Delta  C_p(T-T_0)-
T[\Delta S^0+\Delta  C_p\ln(T/T_0)],
\] 
where $\Delta H^0$, $\Delta S^0$ are respectively the 
enthalpy and entropy of transfer at temperature $T_0=298K$, 
and $\Delta  C_p$ is the  heat capacity (assumed
constant with respect to $T$).

Adopting values collated by Dill {\it et al.}\cite{Dill} $\Delta H^0=0$, 
$\Delta S^0=-6.7$ cal/K/mol, $\Delta  C_p=55$ cal/K/mol, 
hydrophobic surface tension 
in this formulation increases up to a maximum at 60-80$^0$C, subsequently 
decreasing as a consequence of the high heat capacity.

Using $\gamma(T)$, we can specify an {\it effective} hydrophobic contribution 
to the surface tension at any given globule surface, 
provided we know the hydrophobic topology of the surface.
We will represent hydrophobic surface topology 
by a factor $f_h$ defining the fraction of 
amino-acid residues at the surface which are hydrophobic
as opposed to polar, i.e., the effective hydrophobic surface
tension becomes $\sim f_h\gamma$.

Surface tension at the globule surfaces does not in itself 
determine an effective well-depth $\epsilon$ 
characterising globule-globule contacts casually formed in solution.  
We have first to specify the `casual' area of contact 
$\delta A$ to which it is thermodynamically conjugate.

To this end, we minimize the form
\begin{equation}
\epsilon=-\gamma f_h\delta A + \frac{E}{2\pi d}(\delta A)^2,
\end{equation}
where, following earlier work,\cite{Braun} we interpret 
the phenomenological parameter $E\sim$ 10MPa as an osmotic shear modulus 
governing globule elasticity, with $d$ the globule diameter. 

The result is a temperature-dependent well-depth
\begin{equation}
\epsilon(T)=-\frac{\pi d}{2E}[f_h\gamma(T)]^2
\label{epsi}
\end{equation}
(temperature dependence of $E$ is neglected). 

Taking the amino-acid lengthscale $a$ for
the well width, the full potential is
\begin{eqnarray}
v(r,T)&=&\infty\hspace{.5in}0<r<d\nonumber\\
&&\epsilon(T) \hspace{.4in}d<r<d+a\nonumber\\
&&0\hspace{.5in}r>d+a,
\end{eqnarray}

The square-well fluid is not particularly tractable from an
equation of state perspective.  
However, exact and compact analytical results follow
in the adhesive limit formulated by Baxter,\cite{Baxter,Hansen}
\begin{eqnarray}
v(r)&=&\infty\hspace{1.4in}0<r<d\nonumber\\
&&-k_BT\ln\left[\frac{d+\sigma}{12\tau \sigma}\right] \hspace{.4in}d<r<d+\sigma\nonumber\\
&&0\hspace{1.4in}r>d+\sigma,
\end{eqnarray}
where $\sigma\rightarrow 0$.

Baxter's parameter $\tau^{-1}$ can by regarded 
as a measure of adhesive strength, as is clear from
its relation to the second virial coefficient 
\begin{equation}
\Delta B_2/B_2^{HS}=
3d^{-3}\lim_{\sigma\rightarrow 0}\int_d^{d+\sigma}\left[ 1-\exp(-\beta v(r))\right]r^2dr
=-\tau^{-1}/4,
\end{equation}
where the bare hard-sphere result $B_2^{HS}=2\pi d^3/3$ presents a convenient
reference.

The idea of exploiting the analytical tractability of the adhesive limit
to represent square-well-like systems is a familiar one in the general
colloidal context.
We follow here the equivalence prescription of Regnaut and Ravey,\cite{Regnaut} 
who fix a correspondence between $\tau^{-1}$ and well-depth by
equating the respective second virial coefficients.
To lowest order in $a/d$ we have for the square-well fluid,
\[
\Delta B_2/B_2^{HS}
\simeq - 3(a/d)\left[\exp(-\beta \epsilon)-1\right].
\]
 
Hence,
\begin{equation}
\tau^{-1}(T)=12(a/d)\left[\exp(-\beta \epsilon(T))-1 \right].
\label{tau}
\end{equation}

Regnaut and Ravey also give the analytical 
form of the structure factor for Baxter particles, 
\begin{equation}
S_0= \frac{(1-\varphi)^4 }{\left[ 1+2\varphi-\lambda \varphi
(1-\varphi)\right]^2},
\label{s0}
\end{equation}
where $\varphi =\pi d^3c/6$ is, in our case, the protein volume fraction,
and $\lambda$ is the lower root of
\begin{equation}
\frac{\varphi}{12}\lambda^2-\left(\frac{\varphi}{1-\varphi}+\tau\right)\lambda
+\frac{1+\varphi/2}{(1-\varphi)^2}=0.
\label{lam}
\end{equation}

Dispersion force and double layer electrostatics can be treated
via the $B_2$ predictions of DLVO theory (Muschol and
Rosenberger\cite{Muschol} give details of the calculation), yielding
a final expression
\begin{equation}
(S_0)^{-1}= \frac{\left[ 1+2\varphi-\lambda \varphi
(1-\varphi)\right]^2}{(1-\varphi)^4 } +
 2c\Delta B_2^{\rm DLVO} +O(c^2).
\end{equation}

The double layer contribution might alternatively be replaced by a 
Donnan-like treatment of the electrostatics, as presented recently by 
Warren,\cite{Warren} focusing on the ideal distributional entropy of small 
ions in the solution.  In the absence of salt, the charge per protein, $Q$ say, 
must be  balanced according to the charge neutrality 
constraint by a concentration $cQ$ of counterions, 
generating osmotic pressure
\begin{equation}
v\delta \Pi/k_BT= \varphi Q,
\end{equation}
where $v=\pi d^3/6$ is the protein specific volume.

Chemical equilibrium with a salt reservoir 
at concentration $c_s$ leads to the modification
\begin{equation}
v\delta \Pi/k_BT= \sqrt{(\varphi Q)^2+ (2vc_s)^2}.
\end{equation}

Hence, via the fluctuation-dissipation theorem  
$S_0= k_BT (\partial\Pi/\partial c)^{-1}$, and combining with 
Eqn (\ref{s0}), we obtain for charged globules
in the presence of salt 
\begin{equation}
(S_0)^{-1}= \frac{\left[ 1+2\varphi-\lambda \varphi
(1-\varphi)\right]^2}{(1-\varphi)^4 } +
Q\left[1+(2c_s/\,cQ)^2\right]^{-1/2} +
 2c\Delta B_2^{\rm Hamaker} +O(c^2),
\label{mores}
\end{equation}
where $\Delta B_2^{\rm Hamaker}$ is the dispersion force contribution.
Note that the Donnan contribution to $B_2$
follows by comparing the low concentration limit  of this expression
with  $(S_0)^{-1}\sim 1+ 2B_2c $, yielding\cite{Warren} 
$\Delta B_2=Q^2/\, 4c_s$.

This completes our main objective. However, some interesting features in 
respect of the phase diagram deserve comment.
Firstly, there exists for the Baxter system a locus
in $(\varphi,\tau)$-space along which 
the structure factor of Eqn (\ref{s0}) diverges,
tantamount to a phase transition spinodal (diverging
compressibility).
This may or may not underpin a liquid-liquid phase separation 
in the protein solution as represented by Eqn (\ref{mores}),
depending on whether $\lambda$ of the mapping becomes
sufficiently large with respect to the electrostatic term.

Secondly, the `connected cluster' structural view of the Baxter 
fluid reveals a percolation transition, at which the mean cluster size 
$N$ diverges. Cluster size is
formally related to a pair-connectedness 
function $P(r)$ analogous to the usual pair distribution function $g(r)$, 
\begin{equation}
N=1+c\int P(r)d{\bf r},
\end{equation}
where  $c^2P({\bf r}_1,{\bf r}_2)d{\bf r}_1d{\bf r}_2$ 
is the probability of finding connected particles 
in volume elements $d{\bf r}_1$ and $d{\bf r}_2$ simultaneously. 
By recasting Baxter's original $g(r)$ method in connectivity language, Chiew 
and Glandt\cite{Chiew} solve for $P(r)$, yielding
\begin{equation}
N=1/(1-\lambda\varphi)^2
\label{S}
\end{equation}

The percolation transition, i.e. the locus $\lambda\varphi\rightarrow 1$,
is not strongly manifest in extensive thermodynamical quantities,  
there being no free energy singularity. 
However, the qualitative trend exhibited via the mapping into
$(\varphi,T)$-space, percolation 
at lower volume fraction with increasing temperature (since $\lambda$
increases with hydrophobic surface tension $\gamma$), 
is reminiscent of the gelation curve observed experimentally in sickle 
cell hemoglobin solutions.\cite{Eaton}
This resemblance should be qualified by noting that the 
experimentally referred to `gel phase' of HbS 
globules comprises an entangled mesh of multi-stranded polymeric assemblages,
somewhat removed from  the percolation perspective.

It is  useful also to stress that clustering is understood here
in the Hill sense\cite{Hill} implicit to the structure of 
isotropically interacting simple liquids - transient `ordinary liquid clusters' 
of this type are to be contrasted with transient `open clusters' 
characteristic of  systems in which intermolecular 
bonding is orientationally constrained (hydrogen bonding in
liquid water is the classic example).
Open cluster formation, and the attendant prospect of further critical points,
\cite{Tanaka} are of course entirely feasible in the case of 
globular proteins having patch-like hydrophobic surface topology
(as opposed to the uniform topology assumed here via our parameter $f_h$), 
presenting a possible avenue for future investigation.
\medskip

In summary, our main result is Eqn (\ref{mores}) for the
static structure factor, in conjunction with
the mapping via Eqn (\ref{tau}) of a hydrophobic square-well interaction 
specified by Eqn (\ref{epsi}).
As a basis for interpreting experimental light 
scattering from globular protein solutions,  this result is
quantitative to the extent that it takes
into account (i) dispersion force interactions,
(ii) Donnan electrostatics, and (iii) the effective globule-globule 
adhesive interaction which originates from hydrophobicity on the 
amino-acid lengthscale.

Of the unspecified parameters $g$, $f_h$, $E$ introduced into the 
structure factor by the hydrophobic contribution, 
a simple geometrical argument should suffice in order to fix $g$,
while $f_h$ could be obtained in a first approximation from
the sequence information for a given globule, or
directly from tertiary structural data. Insofar as we have less insight into
the phenomenological elastic modulus $E$, a scattering experiment might
focus on estimating this quantity.

Finally, a liquid-liquid spinodal and a percolation line are implicit to 
the general approach.  Although such features are expected to fall more or less 
within the solid-fluid miscibility gap,\cite{Muschol2} they are of 
practical concern to crystallographers and disease pathologists,
in their capacity as metastable phenomena capable of breaking up
the dispersed sol phase of globules, or obstructing crystallization.

\end{document}